# Transition routes of electrokinetic flow in a divergent microchannel with bending walls

Yanxia Shi [1], Ming Zeng [1], Haoxin Bai [1], Shuangshuang Meng [1], Chen Zhang [1], Xiaoqiang Feng [1], Ce Zhang [1], Kaige Wang [1] and Wei Zhao [*,1]

1 State Key Laboratory of Photon-Technology in Western China Energy, International Collaborative Center on Photoelectric Technology and Nano Functional Materials, Institute of Photonics & Photon Technology, Northwest University, Xi'an 710127, China

* Correspondence: zwbayern@nwu.edu.cn (W.Z.)

**ABSTRACT:** Electrokinetic flow can be generated as a highly coupled phenomenon among velocity field, electric conductivity field and electric field. It can exhibit different responses to AC electric fields in different frequency regimes, according to different instability/receptivity mechanisms. In this investigation, by both flow visualization and single-point laser-induced fluorescence (LIF) method, the response of AC electrokinetic flow and the transition routes towards chaos and turbulence have been experimentally investigated. It is found, when the AC frequency $f_f < 30$ Hz, the interface responds at both the neutral frequency of the basic flow and the AC frequency. However, when $f_f \geq 30$ Hz, the interface responds only at the neutral frequency of the basic flow. Both periodic doubling and subcritical bifurcations have been observed in the transition of AC electrokinetic flow. We hope the current investigation can promote our current understanding on the ultrafast transition process of electrokinetic flow from laminar state to turbulence.

Microfluidic systems use microscale channels to process, probe and manipulate small-volume fluids and samples. In recent years, microfluidics has been rapidly developed as a breakthrough technology, which is widely used in various fields, e.g. chemical industry [1], biochemistry [2] and medical field [3], to realize cell culture [4], chemical components test [5] and analysis [6], DNA analysis [7] and liquid mixing [8] etc. As an effective approach, electrokinetic (EK) flow mechanism has been commonly used in the aforementioned research fields and applications.

Previous investigations primarily focus on the electrokinetic instability (EKI) of flow, which mainly describes the development of flow disturbance under the interactions between the electric field and the fluid motion. In 1998, Ramos et al. [9] systematically described the way to drive fluids through electrokinetic mechanisms. They further showed experimental and numerical studies on AC electroosmosis to drive fluid flow at the electrode surface and obtain various fluid velocities at the electrode surface [10-12]. In 2001, Oddy et al. [13] observed EKI in sinusoidal oscillatory electroosmotic channel flow. Relying on this phenomenon, they designed a device to enable rapid mixing of two fluids. Later, Chen et al. [14] observed the EKI of the fluid in a microchannel driven by a DC electric field in an experiment and proposed a physical model for this instability. In 2004, through experiment, theoretical analysis and numerical simulation, Lin et al. [15]. studied the EKI with a conductivity gradient orthogonal to the flow direction and an electric field applied to the flow direction. The results showed when the electric field is over $5 \times 10^4$ V/m, the flow starts to become unstable. Subsequently, in 2006, Posner et al. [16] presented an experimental study of convective EKI in a cross-shaped microchannel, where a low (high) conductivity centre stream flow between two sheath streams of high (low) conductivity under an applied electric field. According to EKI, when the conductivity of the sheath stream is higher than the conductivity of the centre stream, the flow shape exhibits a "pearl-necklace" type structure. In contrast, when the conductivity of the sheath stream is lower than the conductivity of the centre stream, the flow shape exhibits a "wavy" structure. The two structures of flow patterns have also been confirmed in follow-up experimental and numerical simulation research [17,18]. Later, Dubey et al. [19] elucidated the underline mechanisms of wavy and pearl-necklace type structures. When a single interface of electric conductivity gradient is present, an inclined interface can be generated between the high and low electric conductivity fluids. In the limit of high electric Rayleigh numbers, the free charge accumulates at the slanted interfaces, where the net charge ($\rho_e$) is coupled with the electric field to drive the fluid motion. While two conductivity gradient interfaces coexisted, the slanted electric conductivity perturbations can be either in-phase or out-of-phase. For the former, the flow pattern is wavy, and for the latter, the flow pattern is pearl-necklace type structures. In a follow-up study [20], based on the temporal power spectrum analysis of the experimental data, the increase of the electric field first leads to the onset of EKI, and when the field is further increased, chaotic flow can be generated in the microchannel.

Although the investigations above have shown that the flow could evolve into chaotic flow, no direct clues have demonstrated that the unstable EK flow confined in a microchannel can be developed into turbulence. Compared to laminar and chaotic flow, turbulence is a more effective way of enhancing mixing and fluid transport. In macroflow, turbulence can be easily generated. However, in microfluidic systems, due to the smaller channel size and generally lower flow velocity, the Reynolds number is usually very small and the flow state of the fluid is limited to laminar, not turbulence.

In 2014, Wang et al. [21] designed a sophisticated EK active micromixer and found that turbulence could be achieved in a low Reynolds number microflow based on electrokinetic means. They observed many statistical properties which were believed to exist only in macroscale high Reynolds number turbulence [22,23], such as Kolmogorov -5/3 law of the velocity spectra, Obukhov-Corrsin -5/3 law of the concentration spectra and etc. The EK turbulence leads to a fast mixing of momentum and scalar, especially a mixing length of O(100 µm) [24,25].

However, till now, two key issues are still not clear. One is how the EK flow responds to the AC electric field, the other is how the EK flow evolves from laminar to turbulence. This paper mainly presents an experimental study to show the initially linear EKI from a frequency domain using a single-point laser-induced fluorescence (LIF) method. The evolution of the EK flow, from periodic perturbation due to EKI to random turbulent flow, is also investigated by spectral analysis.

## THEORY

The EK flow in this research is generated according to an electric conductivity gradient driven by an AC electric field. It involves both momentum and scalar transport processes, where the momentum transport process can be described by the Navier-Stokes equation with an electric body force (EBF) term [26], as shown below

$$\rho \left( \frac{\partial \boldsymbol{u}}{\partial t} + \boldsymbol{u} \cdot \nabla \boldsymbol{u} \right) = \nabla p + \eta \nabla^2 \boldsymbol{u} + \boldsymbol{F}_e \quad (1)$$

where $\boldsymbol{u} = u\boldsymbol{x} + v\boldsymbol{y} + w\boldsymbol{z}$ represents the velocity vector. $u$, $v$ and $w$ are the velocity components in the $x$, $y$ and $z$ directions respectively as shown in Figure 1(b). $\boldsymbol{x}$, $\boldsymbol{y}$ and $\boldsymbol{z}$ are unit vectors. $\rho$ is the fluid density, $p$ is the pressure on the fluid, and $\eta$ is the coefficient of dynamic viscosity. $\boldsymbol{F}_e$ is the electric body force density, which is expressed as [9,23]

$$\boldsymbol{F}_e = \rho_e \boldsymbol{E} - \frac{1}{2}(\boldsymbol{E} \cdot \boldsymbol{E})\nabla \varepsilon + \frac{1}{2}\nabla \left[ \rho \boldsymbol{E} \cdot \boldsymbol{E} \left( \frac{\partial \varepsilon}{\partial \rho} \right)_T \right] \quad (2)$$

where $\rho_e = \nabla \cdot (\varepsilon \boldsymbol{E})$. Here, we approximate $\varepsilon$ to be constant, then, the third term of Eq. (2) is ignored. The EBF is tightly associated with electric conductivity ($\sigma$) distribution, which is controlled by a convection-diffusion equation

$$\frac{\partial \sigma}{\partial t} + \boldsymbol{u} \cdot \nabla \sigma = D_\sigma \nabla^2 \sigma \quad (3)$$

where $D_\sigma$ is the effective diffusivity of $\sigma$. Furthermore, the flow structures and scalar transport are investigated by laser induced fluorescence through measuring the fluorescent dye concentration ($c$), thus

$$\frac{\partial c}{\partial t} + \boldsymbol{u} \cdot \nabla c = D_m \nabla^2 c \quad (4)$$

where $D_m$ is the diffusivity of fluorescent dye. It should be noted that, in the transport of the scalars, $\sigma$ is an active scalar if an electric field is applied, while $c$ is a passive one regardless of the electric field.

### Linear response regime

Initially, when the electric field intensity is beyond some critical value, the EK flow becomes unstable. The flow can be approximated to 2D initially with small perturbations. Eqs. (1) and (3) become

$$\rho \left( \frac{\partial u}{\partial t} + u\frac{\partial u}{\partial x} + v\frac{\partial u}{\partial y} \right) = -\frac{\partial p}{\partial x} + \eta \left( \frac{\partial^2}{\partial x^2} + \frac{\partial^2}{\partial y^2} \right)u + F_{e_x} \quad (5)$$

$$\rho \left( \frac{\partial v}{\partial t} + u\frac{\partial v}{\partial x} + v\frac{\partial v}{\partial y} \right) = -\frac{\partial p}{\partial y} + \eta \left( \frac{\partial^2}{\partial x^2} + \frac{\partial^2}{\partial y^2} \right)v + F_{e_y} \quad (6)$$

$$\frac{\partial \sigma}{\partial t} + u\frac{\partial \sigma}{\partial x} + v\frac{\partial \sigma}{\partial y} = D_\sigma \left( \frac{\partial^2}{\partial x^2} + \frac{\partial^2}{\partial y^2} \right)\sigma \quad (7)$$

Let $u = u' + U$, $v = V + v'$, $p = P + p'$, $\sigma = \bar{\sigma} + \sigma'$, with $U$ and $V$ being the mean values of $u$ and $v$ respectively. $P$ is the mean value of $p$. $\bar{\sigma}$ is the mean value of $\sigma$. The prime indicates the fluctuation of the physical quantity. After simple processing, the linear perturbation equations are obtained as

$$\rho \left( \frac{\partial u'}{\partial t} + U\frac{\partial u'}{\partial x} + u'\frac{\partial U}{\partial x} + v'\frac{\partial U}{\partial y} \right) = -\frac{\partial p'}{\partial x} + \eta \left( \frac{\partial^2}{\partial x^2} + \frac{\partial^2}{\partial y^2} \right)u' + F_{e_x} - \overline{F_{e_x}} \quad (8)$$

$$\rho \left( \frac{\partial v'}{\partial t} + u'\frac{\partial V}{\partial x} + U\frac{\partial v'}{\partial x} + V\frac{\partial v'}{\partial y} + v'\frac{\partial V}{\partial y} \right) = -\frac{\partial p'}{\partial y} + \eta \left( \frac{\partial^2}{\partial x^2} + \frac{\partial^2}{\partial y^2} \right)v' + F_{e_y} - \overline{F_{e_y}} \quad (9)$$

$$\frac{\partial \sigma'}{\partial t} + u'\frac{\partial \bar{\sigma}}{\partial x} + U\frac{\partial \sigma'}{\partial x} + v'\frac{\partial \bar{\sigma}}{\partial y} = D_\sigma \left( \frac{\partial^2}{\partial x^2} + \frac{\partial^2}{\partial y^2} \right)\sigma' \quad (10)$$

Here, ‾ indicates ensemble average. After making a series of assumptions, e.g. $\frac{\partial V}{\partial x} = \frac{\partial V}{\partial y} = V = 0$ at the centerline of the mixing chamber, initially $u' \ll v'$, $\frac{\partial u'}{\partial x} = -\frac{\partial v'}{\partial y}$ when considering impressibility, and $p' \sim \rho v'^2$, then the linear perturbation equations adopted primarily at the center of the microchannel are simplified as

$$\rho \left( -U\frac{\partial v'}{\partial y} + v'\frac{\partial U}{\partial y} \right) = F_{e_x} - \overline{F_{e_x}} \quad (11)$$

$$\rho \left( \frac{\partial v'}{\partial t} + U\frac{\partial v'}{\partial x} \right) = \eta \left( \frac{\partial^2}{\partial x^2} + \frac{\partial^2}{\partial y^2} \right)v' + F_{e_y} - \overline{F_{e_y}} \quad (12)$$

$$\frac{\partial \sigma'}{\partial t} + U\frac{\partial \sigma'}{\partial x} + v'\frac{\partial \bar{\sigma}}{\partial y} = D_\sigma \left( \frac{\partial^2}{\partial x^2} + \frac{\partial^2}{\partial y^2} \right)\sigma' \quad (13)$$

Since initially, the conductivity gradient is in the $y$-direction only, $F_{e_x} \approx 0$ and $\overline{F_{e_x}} \approx 0$. $\overline{F_{e_y}}$ is only a function of $(x, y)$. Thus, the property of temporal fluctuations of EBF is only determined by $F_{e_y}$. Furthermore, as Eq. (12) is a linear oscillation equation of $v'$ forced by $F_{e_y}$, the response of $v'$ is proportionally determined by $F_{e_y}$ as well.

For an AC electric field $E_y = E_W e^{i\omega_E t} + E_y'$, the electric charge conservation equation [27] is approximated to

$$\frac{\partial \rho_e}{\partial t} = -\boldsymbol{\nabla} \cdot \sigma \boldsymbol{E} = -\frac{\partial}{\partial y}\sigma E_y = -\frac{\partial}{\partial y}(\sigma' + \bar{\sigma})(E_W e^{i\omega_E t} + E_y') \quad (14)$$

where $E_W e^{i\omega_E t}$ denotes the external AC electric field with an angular frequency of $\omega_E = 2\pi f_f$. $f_f$ is the AC frequency and $E_W$ is the amplitude. $E_y'$ represents the electric field perturbations due to $\sigma'$. In the Fourier space of temporal fluctuations,

$$v' = \int_{-\omega_{vc}}^{\omega_{vc}} F(x, y, \omega) e^{i\omega t} d\omega \quad (15)$$

$$E_y' = \int_{-\omega_{ec}}^{\omega_{ec}} F_E(x, y, \omega) e^{i\omega t} d\omega \quad (16)$$

$$\sigma' = \int_{\omega_{\sigma c}}^{\omega_{\sigma c}} F_\sigma(x, y, \omega) e^{i\omega t} d\omega \quad (17)$$

where $\omega_{vc}$, $\omega_{\sigma c}$, $\omega_{ec}$ are the cut-off frequencies of velocity, electric conductivity and electric field fluctuations respectively. For short, we have $F(\omega) = F(x, y, \omega)$, $F_E(\omega) = F_E(x, y, \omega)$ and $F_\sigma(\omega) = F_\sigma(x, y, \omega)$. Thus, the EBF [9] has the following form in Fourier space

$$F_{e_y} = \frac{1}{2}\text{Re}(\rho_e E_y)$$

$$= \frac{1}{2}\text{Re}\left[ (E_W e^{i\omega_E t} + E_y') \int -\frac{\partial}{\partial y}(\sigma' + \bar{\sigma})(E_W e^{i\omega_E t} + E_y') dt \right]$$

$$= \sum_{k=1}^{8} F_k \tag{18}$$

where

$$F_1 = \frac{1}{2}\text{Re}\left[\frac{E_W}{i\omega_E}\left(-\frac{\partial}{\partial y}\bar{\sigma}E_W\right)e^{2i\omega_E t}\right] \tag{19}$$

$$F_2 = \frac{1}{2}\text{Re}\left[\int_{-\omega_{ec}}^{\omega_{ec}}\left(-\frac{\partial}{\partial y}\bar{\sigma}E_W\right)\frac{F_E(\omega)}{i\omega_E}e^{i(\omega+\omega_E)t}d\omega\right] \tag{20}$$

$$F_3 = \frac{1}{2}\text{Re}\left[\int_{-\omega_{\sigma c}}^{\omega_{\sigma c}}-\frac{E_W}{i(\omega+\omega_E)}\frac{\partial F_\sigma(\omega)E_W}{\partial y}e^{i(\omega+2\omega_E)t}d\omega\right] \tag{21}$$

$$F_4 = \frac{1}{2}\text{Re}\left[\int_{-\omega_{ec}}^{\omega_{ec}}-\frac{E_W}{i\omega}\frac{\partial \bar{\sigma}F_E(\omega)}{\partial y}e^{i(\omega+\omega_E)t}d\omega\right] \tag{22}$$

$$F_5 = \frac{1}{2}\text{Re}\left[E_y'\int -\frac{\partial}{\partial y}(\sigma' E_W)e^{i\omega_E t}dt\right]$$
$$= \frac{1}{2}\text{Re}\left[\int_{-\omega_{\sigma c}}^{\omega_{\sigma c}}\int_{-\omega_{ec}}^{\omega_{ec}}-\frac{1}{i(\omega+\omega_E)}F_E(\omega_1)\frac{\partial F_\sigma(\omega)E_W}{\partial y}\right.$$
$$\left. e^{i(\omega+\omega_1+\omega_E)t}d\omega_1 d\omega\right] \tag{23}$$

$$F_6 = \frac{1}{2}\text{Re}\left[E_W e^{i\omega_E t}\int -\frac{\partial}{\partial y}(\sigma' E_y')dt\right]$$
$$= \frac{1}{2}\text{Re}\left\{e^{i\omega_E t}\int_{-\omega_{\sigma c}}^{\omega_{\sigma c}}\int_{-\omega_{ec}}^{\omega_{ec}}-\frac{E_W}{i(\omega_1+\omega)}\frac{\partial}{\partial y}[F_E(\omega_1)F_\sigma(\omega)]\right.$$
$$\left. e^{i(\omega+\omega_1+\omega_E)t}d\omega_1 d\omega\right\} \tag{24}$$

$$F_7 = \frac{1}{2}\text{Re}\left[E_y'\int -\frac{\partial}{\partial y}(\bar{\sigma}E_y')dt\right]$$
$$= \frac{1}{2}\text{Re}\left[\int_{-\omega_{ec}}^{\omega_{ec}}\int_{-\omega_{ec}}^{\omega_{ec}}-\frac{F_E(\omega_1)}{i\omega}\frac{\partial \bar{\sigma}F_E(\omega)}{\partial y}e^{i(\omega+\omega_1)t}d\omega_1 d\omega\right] \tag{25}$$

$$F_8 = \frac{1}{2}\text{Re}\left[E_y'\int -\frac{\partial}{\partial y}(\sigma' E_y')dt\right] \tag{26}$$

According to Eqs. (15-18), the Eq. (12) for $v'$ component can be further rewritten as

$$\rho\left[\int_{-\omega_{vc}}^{\omega_{vc}}i\omega F(\omega)e^{i\omega t}d\omega + U\frac{\partial}{\partial x}\int_{-\omega_{vc}}^{\omega_{vc}}F(\omega)e^{i\omega t}d\omega\right]$$
$$= \eta\left(\frac{\partial^2}{\partial x^2}+\frac{\partial^2}{\partial y^2}\right)\int_{-\omega_{vc}}^{\omega_{vc}}F(\omega)e^{i\omega t}d\omega + \sum_{k=1}^{8}F_k - \overline{F_{e_y}} \tag{27}$$

Since the velocity fluctuation of EK flow is coupled with $\sigma'$ and $E_y'$, we approximately have $\omega_{vc} \approx \omega_{\sigma c} \approx \omega_{ec}$ for low-frequency and large-scale flow structures. Note, this relation is invalid for fully-developed turbulence. Then, approximating $U$ as constant for simplicity in the center region of the microchannel, for incompressible and uniform fluid, the dispersion relation can be expressed as

$$\int_{-\omega_{vc}}^{\omega_{vc}}\left[i\omega\rho + \rho U\frac{\partial}{\partial x} - \eta\left(\frac{\partial^2}{\partial x^2}+\frac{\partial^2}{\partial y^2}\right)\right]F(\omega)e^{i\omega t}d\omega = \sum_{k=1}^{8}F_k - \overline{F_{e_y}} \tag{28}$$

The fluctuations of flow structures are a direct consequence of the AC electric field. However, according to the relations among $\omega_{vc}$, $\omega_{\sigma c}$, $\omega_{ec}$ and $\omega_E$, the EK flow can be distinguished into two cases.

**(1) if $\omega_E \gg \omega_{vc} \approx \omega_{\sigma c} \approx \omega_{ec}$**, the direct coupling between the applied AC electric field and the fluctuations of flow structures is absent. The EK flow cannot immediately respond to the applied AC electric field and will exhibit perturbations at much lower frequencies. Thus, the terms $F_1$ to $F_6$ are negligible in the driving of $v'$, while $F_7$ and $F_8$ must be dominant, even though these are nonlinear forcing terms.

**(2) if $\omega_E \leq \omega_{vc} \approx \omega_{\sigma c} \approx \omega_{ec}$**, the applied AC electric field and the fluctuations of the flow structures have direct and strong coupling. Similar to a forced pendulum, $v'$ share similar spectra as $F_{e_y}$ in a linear instability/receptivity model.

**Nonlinear response regime**

When the applied AC electric field is sufficiently strong, the flow becomes highly disturbed. The flow evolves from linear response to nonlinear response through two routes. One is that the nonlinear forcing terms, including $F_5$ to $F_8$, can introduce multifrequency influence into $F(\omega)$ directly, and thus $F_\sigma(\omega)$ and $F_E(\omega)$. The other is through the interactions of $u'$, $v'$ and $\sigma'$ by their nonlinear terms, as can be seen in the control equations

$$\left(\frac{\partial u'}{\partial t} + U\frac{\partial U}{\partial x} + U\frac{\partial u'}{\partial x} + u'\frac{\partial U}{\partial x} + u'\frac{\partial u'}{\partial x} + v'\frac{\partial U}{\partial y} + v'\frac{\partial u'}{\partial y} + V\frac{\partial U}{\partial y} + V\frac{\partial u'}{\partial y}\right)$$
$$= -\frac{\partial P}{\partial x} - \frac{\partial p'}{\partial x} + \eta\left(\frac{\partial^2}{\partial x^2}+\frac{\partial^2}{\partial y^2}\right)(u'+U) + F_{e_x} \tag{29}$$

$$\rho\left(\frac{\partial v'}{\partial t} + U\frac{\partial V}{\partial x} + u'\frac{\partial V}{\partial x} + U\frac{\partial v'}{\partial x} + u'\frac{\partial v'}{\partial x} + V\frac{\partial V}{\partial y} + v'\frac{\partial V}{\partial y} + V\frac{\partial V}{\partial y} + v'\frac{\partial v'}{\partial y}\right)$$
$$= -\frac{\partial P}{\partial y} - \frac{\partial p'}{\partial y} + \eta\left(\frac{\partial^2}{\partial x^2}+\frac{\partial^2}{\partial y^2}\right)(v'+V) + F_{e_y} \tag{30}$$

$$\frac{\partial \sigma'}{\partial t} + U\frac{\partial \bar{\sigma}}{\partial x} + u'\frac{\partial \bar{\sigma}}{\partial x} + U\frac{\partial \sigma'}{\partial x} + u'\frac{\partial \sigma'}{\partial x} + v'\frac{\partial \bar{\sigma}}{\partial y} + V\frac{\partial \bar{\sigma}}{\partial y} + v'\frac{\partial \sigma'}{\partial y} + V\frac{\partial \sigma'}{\partial y} = D_\sigma\left(\frac{\partial^2 \bar{\sigma}}{\partial x^2} + \frac{\partial^2 \sigma'}{\partial x^2} + \frac{\partial^2 \bar{\sigma}}{\partial y^2} + \frac{\partial^2 \sigma'}{\partial y^2}\right) \tag{31}$$

When the electric field is sufficiently high, strong $v'$ is induced first, and then, strong $\sigma'$ and $u'$ are obtained. The nonlinear terms, e.g. $v'\frac{\partial u'}{\partial y}$, $v'\frac{\partial v'}{\partial y}$, $v'\frac{\partial \sigma'}{\partial y}$, become dominant in the evolution of $F(\omega)$ and $F_\sigma(\omega)$. In the weakly nonlinear region, the evolution of the velocity power spectrum $S(\omega) = F^*(\omega) \cdot F(\omega)$ among different frequency components behave according to resonant interaction theory [28-30]. The theory shows, even if a single frequency component of $F(\omega)$ is injected into the flow, through the nonlinear terms, the kinetic energy can be transported to other frequency components and may eventually form a wide band of $S(\omega)$ and $S_\sigma(\omega) = F_\sigma^*(\omega) \cdot F_\sigma(\omega)$. However, when the flow is highly disturbed with strong nonlinearity, $S(\omega)$ and $S_\sigma(\omega)$ should be described according to Zhao-Wang model [27].

**Fluctuations of the fluorescent dye concentration**

The flow structures and velocity variation can be straightforwardly revealed through the passive transport of fluorescent dye concentration. The concentration fluctuation can be approximately described by

$$\frac{\partial c'}{\partial t} + u'\frac{\partial \bar{c}}{\partial x} + U\frac{\partial c'}{\partial x} + v'\frac{\partial \bar{c}}{\partial y} + V\frac{\partial c'}{\partial y} = D_m\left(\frac{\partial^2}{\partial x^2}+\frac{\partial^2}{\partial y^2}\right)c' \tag{32}$$

with $c'$ and $\bar{c}$ being the fluctuation and mean values of $c$ respectively. Since initially, $u' \ll v'$ and $V \approx 0$, then

$$\frac{\partial c'}{\partial t} + U\frac{\partial c'}{\partial x} + v'\frac{\partial \bar{c}}{\partial y} = D_m\left(\frac{\partial^2}{\partial x^2}+\frac{\partial^2}{\partial y^2}\right)c' \tag{33}$$

In Fourier space,

$$c' = \int_{\omega_{\sigma c}}^{\omega_{\sigma c}}F_c(x,y,\omega)e^{i\omega t}d\omega \tag{34}$$

Let $F_c(\omega) = F_c(x,y,\omega)$, then, for a uniform $U$, initially the dispersion relation can be approximated as

$$\left[i\omega + U\frac{\partial}{\partial x} + D_m\left(\frac{\partial^2}{\partial x^2}+\frac{\partial^2}{\partial y^2}\right)\right]F_c(\omega) = -\frac{\partial \bar{c}}{\partial y}F(\omega) \tag{35}$$

Approximately, the spectral components of $F_c(\omega)$ and $F(\omega)$ are directly related. One important conclusion is, the peak of $F(\omega)$ is also a peak of $F_c(\omega)$.

**Physical model of the EKI.**

The evolution of the EK flow under increasing electric field intensity can be schematically elucidated as in Figure 1. When no AC electric field, due to the intrinsic instability of the divergent channel flow, the interface fluctuates at its neutral frequency slightly, which can be discovered only by a highly sensitive experimental method (Figure 1(a)). Similar results can also be observed if a low-frequency AC electric field with a small amplitude is applied. The difference is the latter fluctuates at the frequency of the AC electric field. As the electric field is further increased, as shown in Figure 1(b, c), the interface becomes highly disturbed. The fluctuations of velocity and electric conductivity near the interface are significantly enhanced, with more frequency components (especially the harmonic and subharmonic ones) developed by the aforementioned nonlinear mechanisms. Then, the emerged frequency components become stronger under higher electric field intensity, which in turn, leads to the broadening of the spectral bands of both velocity and electric conductivity. The flow experiences a transition to chaotic and turbulent, as diagramed in Figure 1(d).

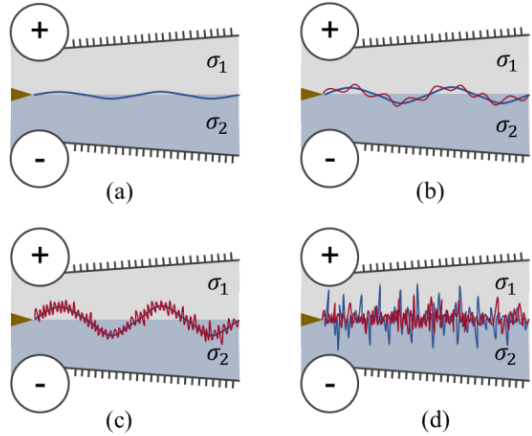

**Figure 1.** Schematic model of the evolution of EK flow with increasing electric field.

## EXPERIMENTAL SETUP

### Optical system

In this research, the electric conductivity fluctuations in a micromixer were investigated with a single-point LIF by measuring fluorescent intensity, through a home-developed confocal microscope, as shown in Figure 2(a). In the system, the excitation beam is generated by a continuous wave (CW) laser of 405 nm wavelength (MDL-III-405-500, CNI). In order to improve the quality of the laser beam, the beam first passes through a spatial light filter (SLF, SFB-16DM, OptoSigma), then collimate with a lens. The laser beam is reflected by a dichroic mirror (Di03-R405/532-t1-25×36, Semrock) and focused into the microchannel to excite fluorescent dye by an Olympus 100× NA 1.4 oil immersion objective lens. The fluorescent signal passes through the dichroic mirror, filtered by a bandpass filter (470/10 nm, OptoSigma), and then, is focused into a multimode optical fiber (25 μm in diameter, 400-550 nm band, M67L01, Thorlabs). At last, the fluorescent signal was detected by a single photon counter (Hamamatsu H7421). The spatial resolution of the confocal microscope is ~180 nm [31].

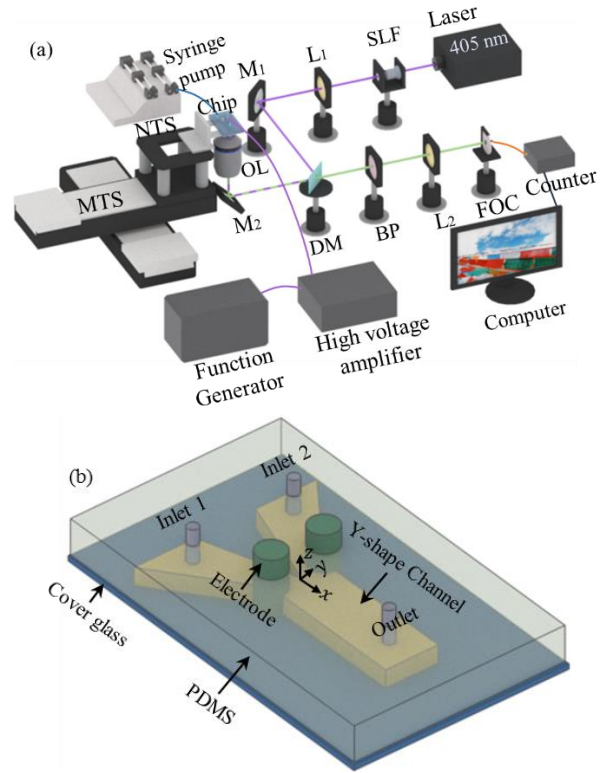

**Figure 2**. Schematic diagram of the experimental system. (a) Schematic of the home-developed confocal microscope, including spatial light filter (*SLF*), lenses (*L*), mirror (*M₁*), dichroic mirror (*DM*), mirror (*M₂*), objective lens (*OL*), bandpass filter (*BPM*), nanocube piezostage (*NTC*, P-562.3CD, PI), translation stage (*MTS*, M-521.DG, PI), (b) Schematic diagram of the micromixer chip.

### Solution Preparation

In the experiment, we used coumarin 102 (C102, Sigma Aldrich) as the fluorescent agent. Its excitation peak locates at 390 nm and the emission peak locates at 470 nm. In the LIF experiments, to avoid the influence of photobleaching on the measurement of fluorescent concentration. Both the excitation beam power and the concentration of C102 are ultra-low. C102 has a concentration of 0.5 μM. During experiments, 3 mL of deionized water and 7 mL of aqueous alcohol (concentration > 99.5%) were mixed to dissolve 2.5 mg C102 powder and prepare a 1 mM C102 fluorescent solution. Then, 20 μL of the 1 mM C102 fluorescent solution was further diluted with 39.98 mL of deionized water to obtain a C102 fluorescent solution at a concentration of 0.5 μM. In the experiment, we keep the electric conductivity ratio between the water solution and C102 fluorescent solution at 1:5250 (water solution with $\sigma_1 = 0.4$ μS cm$^{-1}$ and C102 fluorescent solution with $\sigma_2 = 2100$ μS cm$^{-1}$). The conductivity of the solution is changed by adding phosphate buffered saline (PBS) (HyClone, SH30256.01, USA).

### Fabrication of the micromixer chip

Previous investigations [32] indicate a mixing chamber with slightly divergent side walls can significantly enhance mixing, since such type flow (also known as Jeffery-Hamel flow) is more unstable than that in parallel-plate channel. Nevertheless, in our previous research, the side walls were fabricated directly with two Pt plates. This can cause difficulties in the industrial production of the EK micromixer chip and lead to a waste of expensive metal.

In this investigation, the micromixer was fabricated by soft lithography. The schematic of the chip is shown in Figure 2(b). The micromixer has a splitter plate with a sharp trailing edge at the inlet of the mixing chamber, which is 9 mm long ($l$). Two Pt electrodes of 1 mm diameter are symmetrically assembled at the inlet. The distance between the two electrodes, which is the initial width ($w$) of the mixing chamber, is 650 μm. The electrodes naturally form a divergent mixing chamber with bending surfaces. Thus, the basic flow is also unstable and sensitive to external disturbance. At the downstream of the electrodes, the PDMS walls also have a sustained divergent angle of 5°.

The fabrication process of the micromixer chip is demonstrated in Figure 3. Firstly, a 50 μm thick layer of SU8-3025 photoresist (Microchem, Westborough, Ma, USA) was spin-coated on the silicon wafer at 1500 rpm for 30 s, and then baked. Since the microchannel was designed to have a height of 100 μm, the above procedure was repeated twice. The coated wafer was exposed to UV light at an exposure intensity of 20 mW/cm² in a mask aligner (MIDAS, MDA4LJ, Korea). After exposure, the wafer was further baked on a hotplate. Then, the unexposed SU8-3025 photoresist was removed by SU-8 developer. The PDMS solution (Momentative, RTV-615, USA) was coated onto the patterned wafer, degassed in a vacuum desiccator (Baiji, PC-3, China), and then cured at 80 °C. After curing, the inlet, outlet and electrode holes were punched with a puncher. Finally, the patterned PDMS and coverslips were placed in a plasma washer (Harrick, PDC-002, USA), bonding to finish the fabrication of the micromixer chip.

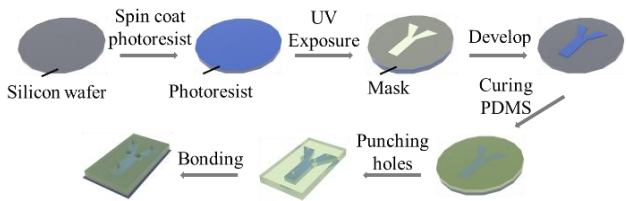

**Figure 3.** Schematic diagram of the fabrication process of the micromixer chip.

## Flow Visualization

The mixing process is monitored by flow visualization, as shown in Figure 4, by an inverted fluorescent microscope (Nexcope, NIB900, China). One solution is deionized (DI) water, and the other is fluorescein sodium salt solution (SIGMA, 46970-100G, Germany). The fluorescent dye is excited by a 473 nm CW laser and emits green fluorescent light around 515 nm. By properly adding PBS buffer solution, the conductivity ratio between DI water and the fluorescent solution is about 1:5000, which is closely consistent with the confocal microscopic measurement. In order to visualize the interface between the two streams clearly with a sufficiently high signal-to-noise ratio, the concentration of fluorescein sodium salt solution is 500 μmol/L. The fluorescence is captured by a SCMOS camera (PCO edge 4.2LT, Germany). The exposure time of the camera is 1 ms.

## EK flow system

A pressure-driven basic flow is provided by a dual-channel syringe pump (HARVARD Apparatus PUMP 33). The fluorescent solution and DI water are pumped into the microchannel through two inlets. The flow rates of both streams are 3 μL/min each. To disturb the flow field electrokinetically, AC electric fields with different frequencies ($f_f$) have been applied by an arbitrary function generator (Tektronix, AFG3102C, USA) and a high-voltage amplifier (Trek, PZD700A, USA). The actual output from the high-voltage amplifier is further monitored by an oscilloscope (Tektronix, MSO2022B, USA).

# EXPERIMENTAL RESULTS

## Flow Visualization

In the experiment, the fluorescein sodium salt solution emits fluorescence under excitation, while DI water does not. When no AC voltage is applied (Figure 4(a)), the flow is steady and stratified, with the concentration interface between the two streams clearly observed. Although the interface is slightly biased to the DI water due to the fabrication deviation, the interface is stable and no distinguishable perturbations were observed from the flow visualization, except an external disturbance (e.g. vibration from the syringe pump or a knock) was applied. In Figure 4(b), when the applied electric field is $f_f = 130$ kHz with $V = 24.2$ V$_{p-p}$, i.e. $E_W = V/2w = 1.86 \times 10^4$ V/m ($V$ is the applied peak-to-peak voltage between two electrodes), it can be seen that the interface is perturbated, indicating the EK flow is no longer steady. As $E_W$ is further increased to $2.74 \times 10^4$ (Figure 4(c)), the interface becomes rolling up and forms an entrainment of the fluids. However, in the current case, the flow has limited structures primarily on large scales. As shown in Figures 4(d) and (e), accompanied by the increasing $E_W$, the perturbation of the flow is significantly enhanced, with a larger spreading rate. The mixing is enhanced as well with small scale flow structures rapidly developed. When $E_W = 8.12 \times 10^4$ V/m, the EK flow exhibits an ultrafast mixing with fine structures, as shown in Figure 4(f). The evolution of the mixing process is consistent with previous studies [21,25] in flat electrodes scheme, indicating the changing of micromixer structure won't intrinsically affect the mixing enhancement through EK turbulence.

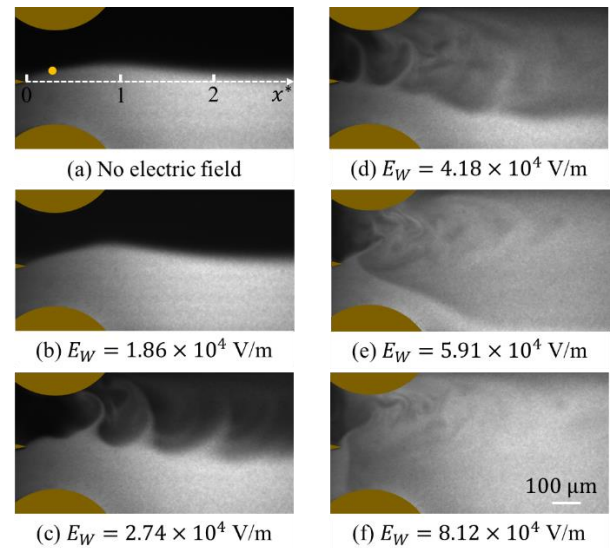

(a) No electric field  (d) $E_W = 4.18 \times 10^4$ V/m

(b) $E_W = 1.86 \times 10^4$ V/m  (e) $E_W = 5.91 \times 10^4$ V/m

(c) $E_W = 2.74 \times 10^4$ V/m  (f) $E_W = 8.12 \times 10^4$ V/m

**Figure 4.** Flow visualization of the flow structures. The conductivity ratio of the two streams is 1:5000. The exposure time of the camera is 1 ms. The flow rate is 3 μL/min for each stream. AC frequency of 130 kHz was applied in the $w = 650$ μm wide micromixer. (a) Without forcing. The yellow dot is the detection position in the following LIF measurements. It locates at $x^* = 0.23$ from the trailing edge. (b) $E_W = 1.86 \times 10^4$ V/m. (c) $E_W = 2.74 \times 10^4$ V/m. (d) $E_W = 4.18 \times 10^4$ V/m. (e) $E_W = 5.91 \times 10^4$ V/m. (f) $E_W = 8.12 \times 10^4$ V/m.

**Single-point LIF for concentration measurement**

At the inlet of the micromixer, a sharp interface is generated between the two streams. The width of the interface can be estimated as $w_{in} = \sqrt{4D_m t} = \sqrt{4D_m x/U_m}$, where $U_m$ is the mean flow velocity at the centerline of the microchannel. At the position $x^* = x/w = 0.23$ where concentration measurement is conducted, assuming $D_m = 5 \times 10^{-10}$ m$^2$/s, $w_{in}$ is approximately 11.4 µm with a larger gradient of fluorescent concentration. At these positions, a small perturbation of the flow can lead to an amplified fluctuation of fluorescence, as shown in Figure 5. Therefore, by detecting the fluctuations of fluorescence, the onset of EK instability can be sensitively measured. With the single-point LIF system, we can continuously monitor the evolution of EK flow from its onset of EKI at a low electric field to a turbulent state at a high electric field.

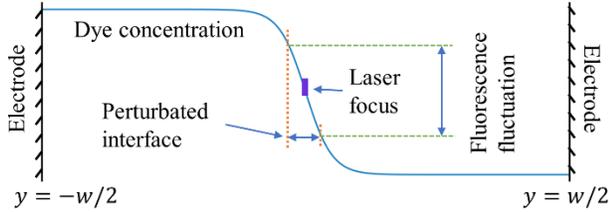

**Figure 5.** Schematic of a small perturbation of the flow lead to an amplified fluctuation of fluorescence.

When using LIF to monitor the fluorescence fluctuations, the influence of photobleaching should be inhibited, otherwise, the fluorescence fluctuations are not only determined by concentration variation, but also contaminated by the velocity fluctuations [22,33-35]. Therefore, we selected a small dye concentration and low excitation beam power to inhibit the influence of photobleaching. The laser power at the pupil is only 0.3 mW and the concentration of the C102 solution is only 0.5 µM. As shown in Figure 6, the fluorescent intensity ($I_f$) is nearly flat with the flow velocity, indicating irrelevance between the fluorescence and flow velocity. The fluorescence is only determined by the concentration of fluorescent dye.

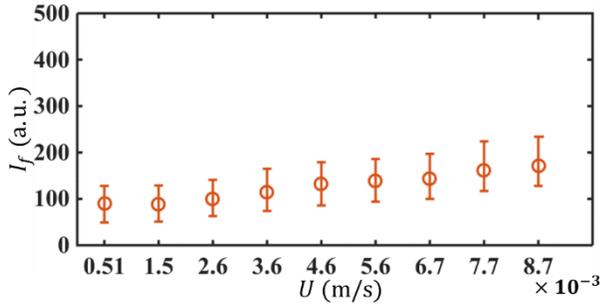

**Figure 6.** Fluorescence intensity of 0.5 µM Coumarin 102 solution under different bulk flow velocity $U$. The pupil laser power is 0.3 mW.

**Time series of fluorescence fluctuations**

Typical concentration fluctuations $c'^*$ ($= c'/c_{rms}$, where $c_{rms} = \sqrt{\overline{c'^2}}$) reflected by the fluorescence fluctuations in the flow with and without AC electric fields are plotted in Figure 7.

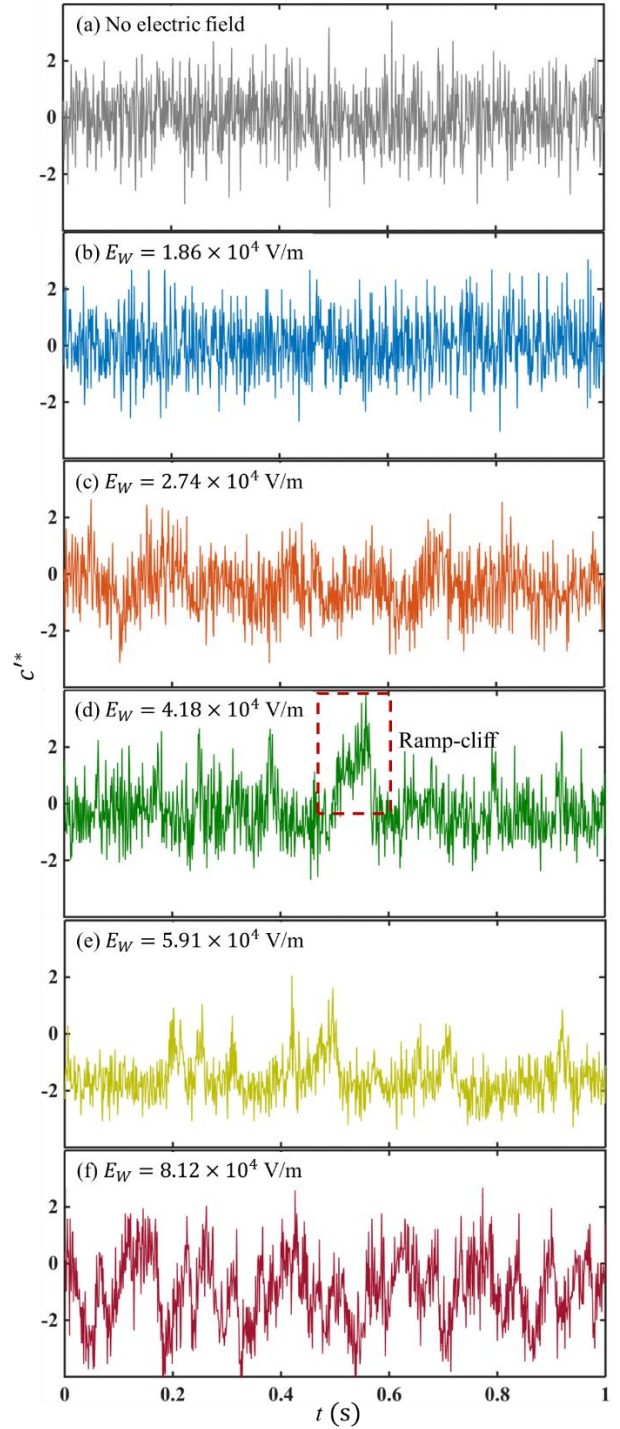

**Figure 7.** Time series of $c'^*$ under different $E_W$ at $Q = 3$ µL/min and $f_f = 130$ kHz. (a) withou forcing, (b) $E_W = 1.86 \times 10^4$ V/m, (c) $E_W = 2.74 \times 10^4$ V/m, (d) $E_W = 4.18 \times 10^4$ V/m, (e) $E_W = 5.91 \times 10^4$ V/m, and (f) $E_W = 8.12 \times 10^4$ V/m.

For the flow without being disturbed by the electric field, as shown in Figure 7(a), $c'^*$ exhibits random fluctuations which are dominated by noise. No flow structures can be distinguished from the time series. Similar results can also be found in Figure 7(b), where a small $E_W$ is applied. As $E_W$ is further increased to $E_W = 2.74 \times 10^4$ V/m, the time series exhibits periodic fluctu-ations (Figure 7(c)) indicating periodic concentration fluctuations in the flow field. This is also consistent with the flow visualization in Figure 4(c).

When $E_W$ is further increased, e.g. at $4.18 \times 10^4$ V/m (Figure 7(d)) and $5.91 \times 10^4$ V/m (Figure 7(e)), $c'^*$ exhibits not only periodicity, but also intermittency, showing a ramp-cliff structure [36], a feature of turbulence from the concentration field. If we keep increasing the electric field to $E_W = 8.12 \times 10^4$ V/m, strong perturbation of $c'^*$ with multiple frequency components can be directly observed from the time series. Relative to Figure 7(a), which is also typical random fluctuations primarily at small scales, the turbulent flow state in Figure 7(f) has more structures with different scales.

The evolution of EK flow with increasing electric field leads to an augmentation of mixing and a transition of flow. This can be seen from the plot of $c_{rms} \sim E_W$ in Figure 8. As $E_W$ is increased, $c_{rms}$ increases first and then decreases. At $E_W = 5.91 \times 10^4$ V/m, a maximum $c_{rms}$ is reached. When $E_W < 5.91 \times 10^4$ V/m, a stronger mixing is reached at a larger $E_W$, indicating stronger flow perturbation and accompanied mixing due to the EK mechanism. The result can also be found in Figure 4. However, when $E_W > 5.91 \times 10^4$ V/m, the declining $c_{rms}$ with $E_w$ indicating the mixing is so fast at sufficiently large $E_w$ that a uniform mixing can be realized adjacent to the trailing edge. This is why $c_{rms}$ becomes smaller. Accordingly, we can deduce the flow is highly disturbed with intense velocity fluctuations. A turbulent state is indirectly unfolded.

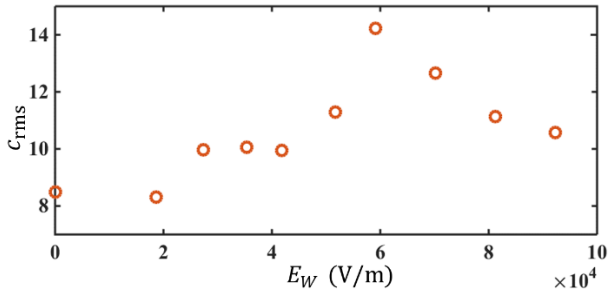

**Figure 8.** $c_{rms}$ vs $E_w$ at $Q = 3$ μL/min and $f_f = 130$ kHz.

**Neutral frequency of basic flow instability**

To reveal the presence of instability at the interface of the two fluids in the micromixer, the frequency responses of the flow with and without AC electric fields are investigated through the power spectrum ($S(f)$) calculated from the time traces of $I'_f$.

Figure 9(a) shows the $S(f)$ without AC electric field, to indicate the possible influence of basic flow instability. Although the bulk flow Reynolds number ($R_e = Ud/\nu$, where $U$, $d$ and $\nu$ are the bulk flow velocity, the hydraulic diameter and the kinematic viscosity) is below 0.33, frequency peaks can always be observed from the spectra, indicating an intrinsically unstable model of the flow. The neutral frequency ($f_n$) of the unstable flow depends strictly on the flow rate, as shown in Figure 9(b). The $f_n$ are all linearly changed with the flow rate $Q$ and uniquely fall on a curve, i.e. $f_n \sim Q$. The results are consistent with that in low Reynolds number wake after sharp trailing edge in macroscale flows [37], where the neutral frequency is approximately $f_n \sim U$. Also, changing the electric conductivity ratio to a much smaller value won't affect the unstable frequency plot.

One interesting observation is, when an AC electric field with $E_W = 2.74 \times 10^4$ V/m and $f_f = 130$ kHz was applied, the response frequency ($f_r$) of the EK flow is exactly the same as $f_n$, for both electric conductivity ratios of 1:2 and 1:5250 (Figure 9(b)). In other words, the EK flow at a high AC frequency doesn't lead to a new instability other than the basic flow.

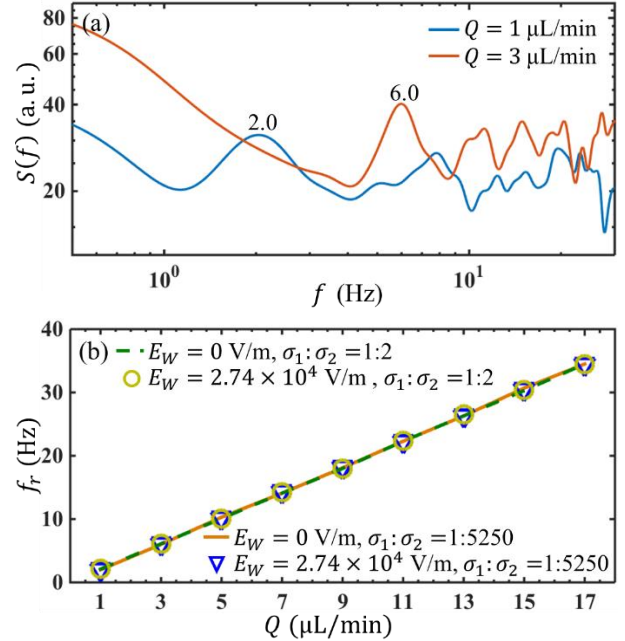

**Figure 9.** (a) $S(f)$ of the unforced flow at different flow rates and conductivity ratio is 1:5250. (b) $f_r$ vs $Q$ at different $E_W$ for conductivity ratios of 1:2 and 1:5250, where $f_f = 130$ kHz.

## $S(f)$ under different $f_f$

### *Initial stage*

To further reveal the frequency response of the EK flow, a broad AC frequency range, from 3 Hz to 130 kHz, has been applied. It is found, depending on the frequency response, the AC frequency range can be distinguished into two subranges, corresponding to two different flow instability/receptivity regimes.

(1) As shown in Figure 10, when the AC frequency is below 30 Hz, for instance, at $f_f = 4$ Hz, there are two spectral peaks observed. One is at the neutral frequency, i.e. $f_n = 6$ Hz, when $Q = 3$ μL/min. The other is at the forcing frequency, i.e. 4 Hz. Since the applied AC voltage is only $V = 8$ V$_{p-p}$, the EK flow fluctuates linearly to the external electric field, forming a forced oscillation system. In this case, the peak of $S(f)$ locates at $f_f$, indicating the concentration fluctuation (which is also the fluctuation of electric conductivity) has $f_r = f_f$. Since the peak of $S(f)$ is also corresponding to the peak of velocity power spectra according to Eq. (35), the flow is perturbed at the AC frequency as well. This observation is interesting since it can only be achieved in two cases. One is that $\rho_e$ still keeps unchanged at a low frequency. The other is really "tricky" that, in $E_W e^{i\omega_E t}$, $\sigma'$ and $E'_y$, two of them have exactly inverse phases, but the same frequencies. Therefore, by the forcing terms $F_2$ to $F_6$, and $F_8$, it is also possible to have $f_r = f_f$.

(2) When the AC frequency is equal to or above 30 Hz, even though a larger electric field $E_W = 2.74 \times 10^4$ V/m has been applied, there is only a single peak observed at the neutral frequency (e.g. $f_n = 6$ Hz). This indicates, on the one hand, when a high-frequency AC electric field was applied, the AC EK instability plays a part to generate flow disturbance indirectly, and relies on the instability of the basic flow. Furthermore, the experimental results also show that at the same AC voltage, the concentration fluctuation is the greatest and most unstable at $f_f = 130$ kHz. This implies, on the other hand, the response of the basic flow to the AC electric field is not the same. The basic flow exhibits a higher receptivity at a higher AC frequency.

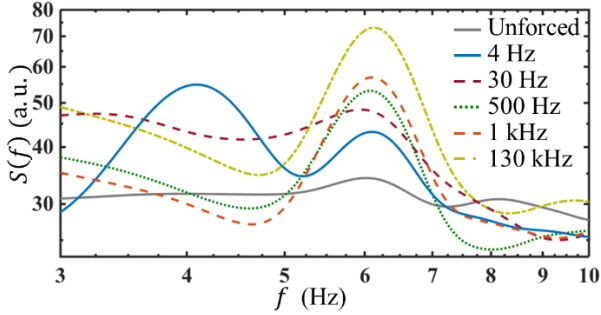

**Figure 10.** Peaks of temporal power spectra of concentration under $E_W = 2.74 \times 10^4$ V/m and different AC frequencies. Here, $Q = 3$ μL/min. The electric conductivity ratio is 1:5250. Note, only at $f_f = 4$ Hz, the applied $E_W = 6.15 \times 10^3$ V/m, to avoid electrolysis.

*Evolution with increasing electric Rayleigh number*

The different receptivity of the flow to AC EBF also leads to a different transitional route to chaos and turbulence. Here, we use $f_f = 4$ Hz as an example to show the evolution of the EK flow with electric Rayleigh number ($Ra_e = 4\frac{\sigma_2-\sigma_1}{\sigma_2+\sigma_1}\varepsilon E_w^2 w_0^2(1-\beta^2)/\rho\nu D_\sigma$) [25]. As shown in Figure 11, when $Ra_e = 3.03 \times 10^4$, $S(f)$ has two peaks at $f_r = f_f$ and $f_r = f_n$ respectively. As $Ra_e$ is increased to $Ra_e = Ra_{ec} = 9.27 \times 10^4$, a second harmonic frequency of $f_f$ is observed, indicating the onset of nonlinearity in the evolution of EK flow, with $Ra_{ec}$ being the critical electric Rayleigh number. Accordingly, linear instability theory becomes inaccurate when $Ra_e \geq Ra_{ec}$.

The peak magnitudes of $S(f)$ increase further with $Ra_e$ till $1.53 \times 10^5$, where $S(f)$ exhibits more peaks at the harmonic frequencies of $f_f$, e.g. at $2f_f = 8$ Hz and $3f_f = 12$ Hz as well. In the following, $S(f)$ experiences a rapid development for both the magnitudes and spectral ranges, as $Ra_e$ is increased to $1.89 \times 10^5$. The harmonic components of $f_f$ grow rapidly first. Then, the subharmonic components, i.e. $f_f/2$ and $3f_f/2$, grow after the harmonic components (see also Figure 12(b)).

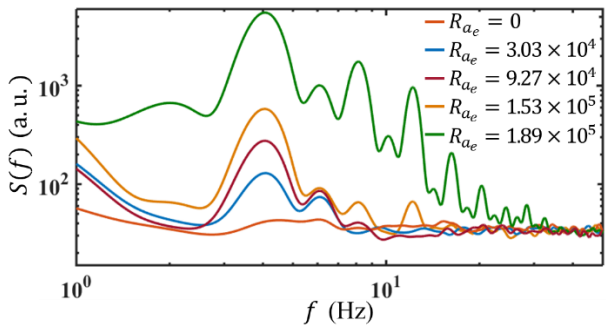

**Figure 11.** Temporal power spectra of concentration under $f_f = 4$ Hz and different electric Rayleigh numbers. Here $Q = 3$ μL/min and the conductivity ratio is 1:5250.

It should be noted, when forcing at low frequencies, the routes to chaos are different at different $f_f$. As shown in Figure 12(a), when $f_f = 3$ Hz, the odd-order harmonic components increase faster and earlier than the even-order counterparts, according to the higher magnitudes at the peaks. While $f_f = 10$ Hz (Figure 12(c)), besides the harmonic components of $f_f$, the peaks also appear at $2f_f/5$, $3f_f/5$ and their combinations.

The evolution of the harmonic and subharmonic frequencies shows that the scalar variance and the corresponding kinetic energy are transported along wavenumbers through both resonant interaction mechanisms (e.g. resonant triads with $\omega_3 = \omega_1 + \omega_2$) [30] along the nonlinear velocity terms and the direct energy injection by nonlinear forcing terms. Both of them lead to the transition of the initially laminar flow to chaotic flow. In the former, the examples can commonly exist from Figure 12. For instance, at $f_f = 10$ Hz, $\omega_1 = 2\pi f_f$ (corresponding to $f/f_f = 1$) and $\omega_2 = \frac{2}{5} \times 2\pi f_f$ (corresponding to $f/f_f = 2/5$), then a peak is observed at $\omega_3 = \frac{7}{5} \times 2\pi f_f$ (corresponding to $f/f_f = 7/5$). At the low AC frequency, a turbulent flow has not been observed, since the applied electric field cannot be sufficiently high. Otherwise, electrolysis can induce air bubbles which severely affect the flow field and internal electric field.

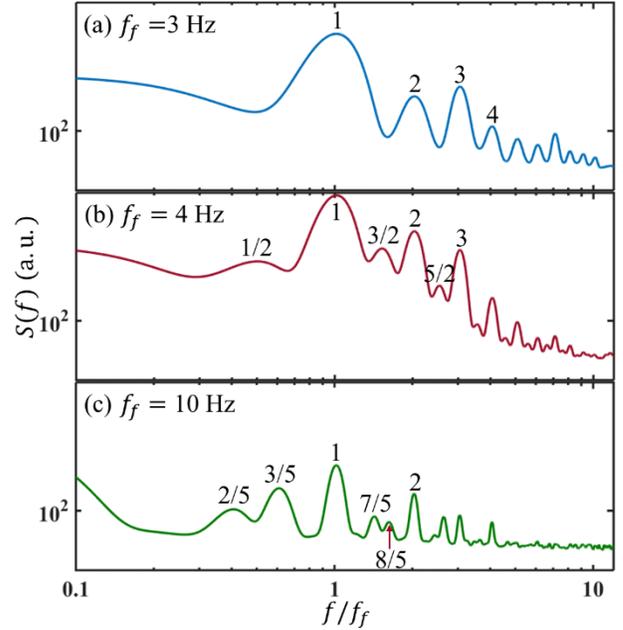

**Figure 12.** Peaks of temporal power spectra of concentration under $V = 35.6$ $V_{p-p}$ at 3 μL/min and different forcing frequencies. (a) $f_f = 3$ Hz. (b) $f_f = 4$ Hz. (c) $f_f = 10$ Hz.

When $f_f$ is equal to or beyond 30 Hz, there is no receptivity at the AC frequency. For instance, under $Q = 3$ μL/min, when unforced, there is a small peak at 6 Hz corresponding to the intrinsically unstable mode of the basic flow, as shown in Figure 13. When forced at $f_f = 130$ kHz and $Ra_e = 2.77 \times 10^5$, the peak of $S(f)$ is still at 6 Hz, with the magnitude clearly enhanced. As $Ra_e$ increases to $Ra_{ec} = 6.00 \times 10^5$, beyond the fundamental frequency (i.e. 6 Hz), $S(f)$ exhibits an additional peak at the second harmonic frequency, i.e. 12 Hz.

When the electric Rayleigh number is further increased to $Ra_e = 1.00 \times 10^6$, the magnitudes of $S(f)$ increase rapidly,

with a much wider spectral band. A slight shift of the fundamental frequency is also observed. The second, third and higher-order harmonics all exhibit large magnitudes, which can be attributed to the nonlinear forcing terms, e.g. $F_8$, and the nonlinear coupling terms, e.g. $v'\frac{\partial v'}{\partial y}$ and $v'\frac{\partial \sigma'}{\partial y}$ etc. Due to the nonliearity, the EK flow exhibits more small-scale and high-frequency structures, which almost form a continuous spectrum. When $Ra_e = 1.40 \times 10^6$, the fundamental frequency continuously shifts towards higher frequency, with increasing bandwidth. As $Ra_e$ is finally increased to $2.14 \times 10^6$, a broad and continuous spectrum with a slope of -1.38 is formed. From the figure, it can be seen the scaling subrange appears first at high frequency, and develops towards both higher and lower frequency regimes. This is consistent with the Zhao-Wang model [27] and supports that EBF is dominant on small scales. Furthermore, from both the flow visualization and $S(f)$, it can be concluded that EK flow is turbulent now.

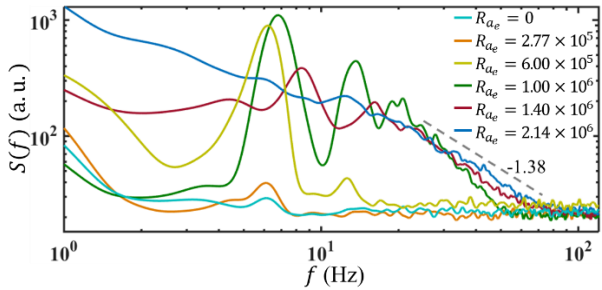

**Figure 13.** Temporal power spectra of concentration under $f_f = 130$ kHz and different electric Rayleigh numbers. Here $Q = 3$ μL/min and the conductivity ratio of the streams is 1:5250.

## DISCUSSION

To this end, the routes of the EK flow toward chaotic and turbulent states have been revealed experimentally, as diagramed in Figure 14. When $Ra_e$ is increased, a series of bifurcations emerge, however, through different routes relying on $f_f$. For instance, in Figure 14(a), when $f_f = 4$ Hz, the nonlinearity of EK flow initiates at $Ra_{ec} = 9.27 \times 10^4$. When $Ra_e \geq 1.89 \times 10^5$, $S(f)$ shows a peak at $f_f/2$ indicating the onset of periodic doubling bifurcation [38]. Then, as can be more clearly explained by resonant interaction theory [28-30], a series of peaks is generated at both the harmonic frequencies of $f_f$ and $f_f/2$, and towards chaos when $Ra_e$ is further increased.

While for the flow driven by $f_f = 130$ kHz, the transition might be attributed to a subcritical bifurcation, according to recent numerical investigations [39] on electroconvection. Its development can also be presented by the resonant interaction theory [28-30] at the weakly nonlinear regime and lead to the fast development of EK flow in Figure 14(b). Nevertheless, a deeper understanding of the transitions of the EK flow can only be reached with a detailed and time-resolved numerical simulation.

## CONCLUSION

In this investigation, the evolution of electrokinetic flow in a microfluidics chip with cylindrical electrodes and divergent walls has been investigated experimentally. It is found that when an AC electric field is applied to the flow, the interface is perturbed with different response frequencies. When $f_f < 30$ Hz, the interface responds at both the neutral frequency of the basic flow and the AC frequency. When $f_f \geq 30$ Hz, the interface responds only at the neutral frequency of the basic flow. The different responses to the AC frequency reveal different instability/receptivity mechanisms of the electrokinetic flow, and accordingly, the different routes to chaos and turbulence. We hope the present work provides an experimental foundation to understand the complex interactions among the velocity field, electric conductivity field and electric field. It could also contribute to the design of novel microfluidic devices.

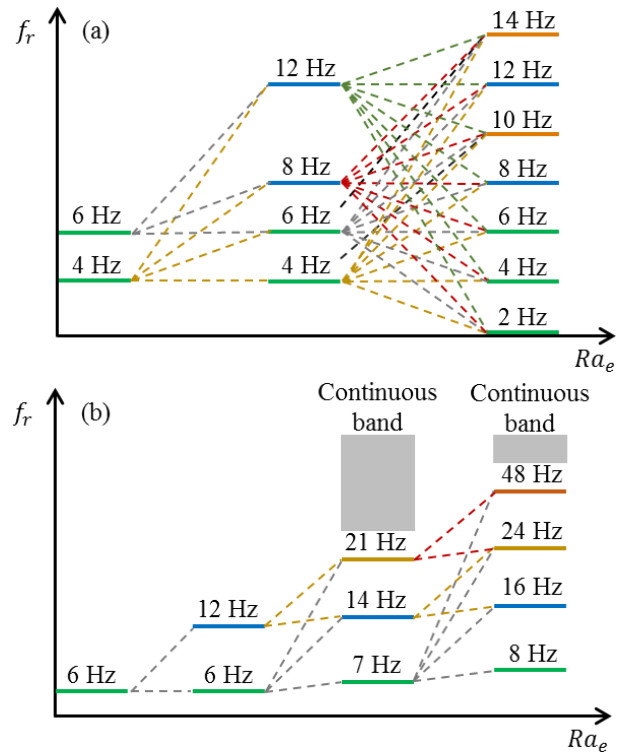

Figure 14. Schematic of $f_r$ vs $Ra_e$, where $Q = 3$ μL/min and the conductivity ratio of the streams is 1:5250. (a) $f_f = 4$ Hz, (b) $f_f = 130$ kHz.